\documentclass[aps,twocolumn,groupedaddress]{revtex4}
\usepackage{amsmath}
\usepackage{latexsym}

\newcommand{\lpl}{$l_{\rm pl}~$} \newcommand{\mpl}{m_{\rm pl}}

\begin{document}
\bibliographystyle{apsrev}

\title{Inflation with a Planck-scale frequency cutoff}

\author{Jens C. Niemeyer} \email[]{j-niemeyer@uchicago.edu}
\affiliation{Max-Planck-Institut f\"ur Astrophysik,
Karl-Schwarzschild-Str. 1, D-85748 Garching, Germany}

\begin{abstract}
The implementation of a Planck-scale high frequency and short
wavelength cutoff in quantum theories on expanding backgrounds may
have potentially nontrivial implications, such as the breaking of
local Lorentz invariance and the existence of a yet unknown mechanism
for the creation of vacuum modes. In scenarios where inflation begins
close to the cutoff scale, these effects could have observable
consequences as trans-Planckian modes are redshifted to cosmological
scales. In close analogy with similar studies of Hawking radiation, a
simple theory of a minimally coupled scalar field in de Sitter space
is studied, with a high frequency cutoff imposed by a nonlinear
dispersion relation. Under certain conditions the model predicts
deviations from the standard inflationary scenario. We  also comment
on the difficulties in generalizing fluid  models of Hawking radiation
to cosmological space-times.
\end{abstract}


\maketitle

\section{Introduction}
\label{intro}

Quantum field theory as we know it is expected to break down near the
Planck scale, \lpl $\sim \mpl^{-1}$  \footnote{We shall use natural
units, $c = \hbar = 1$, throughout this paper so that $\mpl$
represents the Planck mass, energy, frequency, wavenumber, and
momentum.}, where dimensional analysis indicates that gravity becomes
as strong as the standard model forces. At this scale, nature is
believed by many to contain an intrinsic cutoff of the fundamental
degrees of freedom that will ultimately help to cure the divergencies
of quantum field theory and general relativity. The manifestation of
this cutoff in quantum gravity is as yet unknown; it may, for
instance, render the theory discrete or invalidate the notion of
locality as suggested by matrix theory \cite{BFSe97}. It is also
unclear at which energy or  momentum scale the cutoff begins to
noticeably affect the behavior of quantum fields. For simplicity, the
terms ``Planck scale'' and ``cutoff scale'' will be considered
equivalent in this paper, bearing in mind that they might turn out to
be very different concepts. 

Under almost all circumstances, long wavelength phenomena are
sufficiently decoupled from small scales  to hide the short distance
behavior of the fundamental theory. There are two well known
situations where low energy quantum states potentially retain a memory
of their Planckian origin: Hawking radiation and cosmological
expansion (see \cite{J00} for a recent overview of the trans-Planckian
puzzle in both contexts). 

In the original derivation of Hawking radiation, the exponential
redshift of outgoing radiation immediately outside of a  black hole
event horizon requires the existence of a trans-Planckian mode
reservoir \cite{H74a,H74b}. If Planck scale physics does indeed
provide a short distance or high frequency cutoff, where do these
modes come from \cite{J91}?  Beginning with Unruh's sonic black hole
analogy \cite{U81,U95}, whose basic philosophy we will employ in
Sec.~\ref{model}, several papers have addressed this question by
studying linear scalar field theories with nonlinear dispersion
relations  \cite{BMPe95,CJ96,C98,HT00,SS00}, short distance
uncertainty relations \cite{BGLe99}, and discretized spatial degrees
of freedom \cite{CJ98,JM00}. Most of these models produce Hawking
radiation by mode conversion \cite{J96} of ingoing modes into outgoing
ones without affecting the thermal Hawking spectrum, provided that the
cutoff scale is well separated from the black hole horizon scale. In
the lattice model, mode conversion takes place by means of Bloch
oscillations. However, in the present formulation the lattice is
constantly expanding, invalidating the notion of a fixed cutoff length
\cite{JM00}. The nonlinear dispersion models achieve mode conversion
by a reversal of group velocity near the horizon, but they fail to
fully account for the origin of Hawking radiation owing to their time
independence and hence the conservation of Killing frequency
\cite{CJ98}. Thus it seems that in order to formulate a self-contained
model of Hawking radiation that avoids any reference to trans-Plankian
modes one needs to include an explicit mechanism for the creation and
dissipation of modes \cite{J00}. It will be argued below that this
statement is even more relevant in the cosmological context where mode
conversion cannot take place.

In this work, an attempt is made to use a similar approach to the
other setting that involves strong redshifting, i.e.~cosmological
expansion. Apart from the obvious overlap with the black hole problem,
there are at least three different motivations for this inquiry. The
most striking implications of a Planck scale cutoff arise in scenarios
where the inflationary phase lasts sufficiently long to redshift
trans-Planckian wavelengths to cosmological scales as in the simplest
models for chaotic inflation \cite{L84}. This case is especially
interesting since it, at least in principle, allows for observable
Planck scale relics reflected by non-Gaussianity and broken scale
invariance of the distribution of cosmic microwave background
perturbations (Sec.~\ref{implications}). In Sec.~\ref{model}, a simple
model for inflation near the cutoff scale is analyzed, featuring a
modified nonlinear dispersion relation of the form employed in
\cite{U95} that imposes an upper bound on the proper frequency of a
free scalar field on a de Sitter background.

A detailed analysis of the sensitivity of inflation to trans-Planckian
behavior using a very similar approach is described in
Refs.~\cite{MB00,BM00}. Choosing three different types of dispersion
relations and two alternative vacuum states, the authors study the
scale dependence of the predicted perturbation power spectrum, finding
significant deviations from scale invariance under certain
conditions. While our work was done independently, the relation of our
results to those of Refs.~\cite{MB00,BM00} will be highlighted at the
appropriate points.

But even in the more benign case of ordinary power law cosmological
expansion the introduction of a proper frequency or distance cutoff
presents a conceptual challenge. As shown in \cite{W85} and further
discussed in \cite{J00}, the formulation of a Hamiltonian lattice
field theory in an expanding space-time is highly
problematic. Furthermore, a dynamically expanding or contracting
universe with a proper cutoff implies a growing or shrinking Hilbert
space of the effective field theory living in it \cite{J00} and it is
not obvious how this can be realized while preserving unitarity and
local Lorentz invariance on large scales. As always, the final answer
must be deferred to quantum gravity, but it may be useful to take a
phenomenological stance and ask the following question: Can a field
theory in an expanding universe be modified in an ad hoc, but simple
and  well controlled way such that its proper energy and momentum
eigenvalues are bounded but it asymptotically reduces to a standard
relativistic field theory for measurements that involve only large
scales? Just like in similar investigations of the Hawking  effect and
its sonic analogies, we expect to learn something about the
requirements for a full theory by trying (and often by failing) to
answer this question. This program is, of course, beyond the scope of
the present paper but we hope to raise some of the relevant questions
in Sec.~\ref{implications}.

Finally, issues regarding Planck scale effects on the dynamics of
inflation may be even more relevant in the framework of brane
cosmology \cite{RS99} and large extra dimensions \cite{ADD99}, where
the fundamental energy scale might be substantially lower than the
conventional value for $\mpl$, and where the effective Planck scale
felt on our brane is a time variable quantity \cite{ADKe00}. In these
models, it is conceivable that Planck scale physics of the form
discussed in Sec.~\ref{model} has left an imprint on the cosmic
microwave background. 

\section{Some implications of a fixed cutoff in expanding
space-times}
\label{implications}

\subsection{The problem and its possible circumvention}

The trans-Planckian puzzle in the context of inflationary cosmology
was pointed out by Brandenberger \cite{B99}. In many scenarios for
inflation, most notably the chaotic variety, the inflationary phase
commences when the Hubble length $H^{-1}$ is of the order of \lpl and
lasts for $N \sim 10^8$ $e$-foldings \cite{L84,KT90} (only the second
property is relevant at this point, the first will become important in
Sec.~\ref{model}). If a cutoff were present in the theory, metric
perturbations on cosmological scales today would have redshifted from
modes that were not part of our universe when inflation began. To see
this, assume that at some initial time $t_{\rm i}$  the mode expansion
of linear inflaton perturbations in terms of the comoving wavenumber
$k$ is truncated at some high wavenumber $k_0 = a(t_{\rm i}) \mpl$,
where $a(t)$ is the scale factor of the Friedmann-Robertson-Walker
(FRW) line element. Due to the absence of mode interactions in a
linear theory the comoving cutoff wavenumber $k_0$ is conserved,
implying that at the time  $t_{\rm f}$ after  inflation the cutoff
corresponds to a physical scale $l_{\rm c} = a(t_{\rm f}) k_0^{-1} =
e^N$ \lpl. 

Of course, one cannot naively impose a cutoff on an initial spatial
hypersurface at the onset of inflation and evolve the fields without
modifying the theory, as this would lead not only to a suppression of
inflaton fluctuations but of all quantum modes in the universe today
\cite{J91}. Moreover, all of the mechanisms for mode conversion of
ingoing into outgoing modes that salvage Hawking radiation in
truncated theories \cite{U95,BMPe95,CJ96,CJ98} hinge on the spatial
inhomogeneity of the black hole space-time and therefore seem unlikely
to work in a FRW universe where no meaningful separation of
``ingoing'' and ``outgoing'' modes exists. Hence, in order to keep the
cutoff scale pegged to a fixed absolute value, new modes must be added
dynamically as the universe expands or deleted as it collapses. We
will refer to this process as ``mode creation'' or ``mode
dissipation''. 

Since only $\sim 60$ $e$-foldings of inflation are needed to explain
the observed flatness of the universe, the problem can be evaded for
purely practical purposes by demanding that inflation terminated
before the initial cutoff length had crossed the particle horizon
\footnote{This is in contrast with the black hole case where a given
mode  detected by a distant observer was redshifted exponentially in
time since the black hole collapsed.}. While this solution effectively
hides Planck scale physics from cosmological observations, it doesn't
alleviate the need for mode creation and must therefore be considered
incomplete. 
 
In order to clarify the role of ultra high frequencies in the Hawking
effect, Jacobson \cite{J93} demonstrates an alternative derivation
that makes no reference to frequencies above a certain
cutoff. Instead, it imposes a  boundary condition in a timelike region
near the horizon that forces quantum fields to be in their vacuum
state for frequencies larger than the inverse black hole mass but
smaller than the cutoff frequency, where all frequencies are measured
in the frame falling freely into the black hole. One can use a very
similar construction to discuss the dynamics of linear inflaton-driven
metric perturbations in the presence of a cutoff. In the gauge
invariant formalism \cite{M88}, the quantized degree of freedom $\hat
v$ is a linear combination of inflaton and scalar metric perturbations
obeying the equation of motion for a minimally coupled scalar field
with a time dependent mass. At large wavenumbers, $k \gg a H$, its
plane wave mode functions $v_k$  asymptotically behave like those of a
massless scalar field on a de Sitter background, justifying the choice
of the adiabatic massless de Sitter-invariant vacuum for these modes. 

Instead of prescribing the vacuum condition as an initial condition
for integrating the mode equations as in Ref.~\cite{M88}, it can be
re-interpreted as a boundary condition in $k$-space that is imposed at
all times during inflation. In situations where the horizon length is
far removed from the cutoff, $H^{-1} \gg$ \lpl, it is possible to
separate scales between those close to the cutoff and those where the
$v_k$ are still de Sitter-like.  In other words, if the adiabatic
vacuum boundary  condition holds for  $a H \ll k \ll a \mpl$, the
predictions for the amplitude of $|v_k|^2$ at horizon crossing, and
hence for the power spectrum of cosmological perturbations, remain
unaltered. 

Lacking a theory of mode creation, the best we can do to protect
low-energy effective field theories in an expanding universe from
cutoff anomalies is to demand the validity of the above boundary
condition in cases where $H^{-1} \gg$ \lpl is true (which it certainly
is today). We can now ask the following questions: first, is it
possible to isolate some key properties of a mode creation process
that satisfies the boundary condition in this asymptotic regime and
second, once we have constructed such a theory, how do the predictions
for cosmological perturbations change in cases where the scale
separation fails, i.e.~if $H^{-1} \sim$ \lpl? 

Constraining the investigation to theories that obey the asymptotic
boundary condition, it is clear that the effects of a cutoff on the
spectrum of cosmological perturbations should be strongest on those
scales that crossed the horizon when it was smallest and hence closest
to the cutoff scale. As $H^{-1}$ grows during standard inflation,
those scales would now be the largest observational ones.

\subsection{Breaking of local Lorentz invariance}

Perhaps the most prominent implication of a short distance or high
frequency cutoff defined in absolute terms (rather than by
interactions) is a violation of local Lorentz invariance, since a
preferred frame must be specified in which the cutoff scale is to be
measured. Furthermore, local Lorentz non-invariance may be a necessary
ingredient of a fundamental theory in order to avoid a diverging
number of degrees of freedom per unit volume \cite{J91}. 

If our universe is indeed homogeneous and isotropic on large scales, a
preferred frame is already picked out by the FRW cosmic rest frame,
i.e.~the rest frame of the cosmic microwave background radiation. As a
first, simple step we will assume in Sec.~\ref{model} that ``high''
frequencies are defined with respect to the cosmic rest
frame. Ultimately, however, it would be preferable to have a
prescription for the measurement of \lpl that is independent of the
global symmetries of the universe. A possible approach is to use a
dynamical unit timelike vector field to define the preferred frame
\cite{JM00b}.

\subsection{The fluid analogy: dispersion and dissipation}

Looking for guidance in the construction of a phenomenological model
for mode creation we turn to the fluid analogy that proved so useful
for analyzing the effects of a cutoff on Hawking radiation. Unruh
\cite{U81} showed that the quantized sound field of a fluid flow
containing a sonic horizon becomes thermally excited for the same
reason that a black hole emits Hawking radiation. Providing a natural
short distance cutoff due to the breakdown of the continuum
assumption, fluid models represent a well-defined testbed for Hawking
radiation.  

Sound waves with ever decreasing wavelengths propagating in a fluid
first perceive the approach of the molecular scale $l_{\rm m}$ by a
change in the dispersion relation coupling the wavenumber $k$ and the
frequency $\omega$. Typically, the dispersion relation is linear in
the low wavenumber regime and dips over toward smaller frequencies as
$k \to k_0 \sim l_{\rm m}^{-1}$, never exceeding a maximum frequency
$\omega_0$. For studies of Hawking radiation, this  behavior was
mimicked by invoking an artificial  nonlinear dispersion relation in a
linear scalar field theory living on a two-dimensional black hole
space-time \cite{U95,BMPe95,CJ96}. It was found that wave packets
propagated backwards in time are ``reflected'' off the horizon by
virtue of the group velocity dropping below the speed of sound (or
light) as the waves are blueshifted toward the cutoff. The origin of
the outgoing Hawking modes was thus revealed to be exotic ingoing
modes that were ``converted'' into outgoing ones by the spatially
varying group velocity.

In contrast, all modes in an expanding or  collapsing spatially
homogeneous FRW universe are red- or blueshifted at the same rate,
making mode conversion impossible. There is little hope to avoid an
explicit prescription for mode  creation and dissipation if both
energy and momentum are to be bounded. Turning again to fluids for
intuition, here viscosity is responsible for both dispersion and
dissipation, i.e.~the dispersion relation becomes both nonlinear and
complex \cite{V97}\footnote{In order to reduce the problem of the
modified Hawking effect to solving single-mode ODEs numerically
\cite{CJ96} or analytically \cite{C98}, Corley \& Jacobson use a
nonlinear dispersion relation that becomes complex above the cutoff
wavenumber. Its motivation is different from the (inverse) dissipation
framework discussed here, and its validity in the complex regime is
unclear.}. In spite of being manifestly non-unitary, this is a
possible model for a collapsing universe where we want to get rid of
existing modes. Mode creation in an expanding space-time, on the other
hand, would involve inverse dissipation, i.e.~exponential growth of
the wave amplitudes from initial data arbitrarily close, but not
identical, to zero. There is no obvious choice for assigning this
initial data in the very early universe \footnote{The closest analogy
to thermal noise in fluids might be to prescribe some sort of
``space-time noise'' near the Planck scale. Perhaps a phenomenological
approach similar to the one sketched in Refs.~\cite{A99,EMN99} may
turn out useful.}. Perhaps even more  importantly, in order to satisfy
the adiabatic vacuum boundary condition of Sec.~\ref{implications},
the redshifted modes ultimately need to be stabilized at a given
amplitude and phase. This implies either tremendous fine-tuning of the
initial data or nonlinear mode interactions. The latter appears
possible in principle, but finding a viable form for the interaction
that fulfills all the above requirements may be a formidable
challenge. 

The situation is little better for lattice theories on expanding
backgrounds \cite{W85}. It seems to be necessary to dynamically
introduce new lattice points as the universe expands \cite{J00}. Doing
so, one basically faces the same dilemma for assigning field values to
these points as in the inverse dissipation model above. 

The first and simplest choice for analyzing the effects of a Planck
scale cutoff on inflationary perturbations is thus to ignore the
actual creation of the modes and focus only on their frequency
evolution. Since no dissipation is involved, this case is entirely
analogous to the sonic black hole analysis in \cite{U95}. Just like
there, we want to construct a model whose proper frequency is bounded
when the proper wavelength drops below a critical value. Evolving this
mode backward in time, we would see it   oscillate more and more
slowly with respect to the cosmological expansion until it freezes in,
while its wavelength continues to be blueshifted without bound. 

The same general approach to the trans-Planckian problem in
inflationary cosmology was chosen by the authors of
Refs.~\cite{MB00,BM00}. In addition to the asymptotically constant
Unruh dispersion relation described below, two additional forms of
dispersion relation (the sub/superluminal cases of Ref.~\cite{CJ96})
were analyzed in two different vacuum states.

\section{A simple model for inflation near the cutoff scale}
\label{model}

The setting for the model discussed in this section is the slow roll
phase of inflation, driven by a scalar field rolling down a very flat
potential. At the level of simplicity we are seeking it is unnecessary
to specify any details of the potential or the amplitude of the
homogeneous inflaton mode. Keeping in mind that in a more realistic
scenario, the Hubble parameter $H$ is a slowly varying function of
time, it will be assumed constant for our purpose of finding the first
order effects of a nonlinear inflaton dispersion relation. The
cosmological background is thus given by exact de Sitter space. As a
further simplification, the gauge invariant inflaton-metric
perturbation variable $\hat v$ \cite{M88} will be replaced by a free,
massless scalar field mimicking linear inflaton perturbations. This is
a very good approximation for a weakly self-coupled inflaton field
during slow roll \cite{KT90}.

\subsection{Equations of motion}

In order to establish a framework for the model that includes the
proper frequency cutoff, let us first recall the standard theory for a
scalar field in de Sitter space where $\omega$ and $k$ are both
unbounded. A minimally coupled massless scalar field $\phi$ obeys the
equation of motion 
\begin{equation}
\label{covKG}
\Box \phi = 0\,\,,
\end{equation}
where $\Box$ stands for the covariant Laplace-Beltrami operator. In
terms of the conformally flat metric for de Sitter space,
\begin{equation}
\label{conf}
ds^2 = a(\eta)^2 ( d\eta^2 - d{\bf x}^2 ) \,\,,
\end{equation}
with the conformal time $\eta$, the cosmological scale factor $a(\eta)
= - (H \eta)^{-1}$, and the 3-dimensional Euclidian space element
$d{\bf x}$, Eq.~(\ref{covKG}) becomes:
\begin{equation}
\label{dSKG}
\phi'' + 2 \alpha \phi' - \nabla^2 \phi = 0\,\,,
\end{equation}
where $\alpha \equiv a'(\eta)/a(\eta)$, and the prime denotes
differentiation with respect to $\eta$. $\phi$ can be expanded into
comoving plane waves \cite{BD84},
\begin{equation} 
\phi({\bf x},\eta) = (2 \pi)^{-3/2} \int_{\bf k} \left(  a_{\bf k}
u_{\bf k}({\bf x},\eta) +  a_{\bf k}^\dagger u_{\bf k}^\ast({\bf
x},\eta)\right) d{\bf k}
\end{equation}
where
\begin{equation}
u_{\bf k}({\bf x},\eta) = a(\eta)^{-1} \, \chi(\eta) \,  e^{i{\bf k
x}}\,\,,
\end{equation}
and the mode functions $\chi(\eta)$ satisfy ($k=|{\bf k}|$):
\begin{equation}
\label{mode}
\chi'' + \left(k^2 - \frac{a''(\eta)}{a(\eta)}\right)\chi = 0\,\,.
\end{equation}
On subhorizon scales, $k \gg a(\eta) H$, Eq.~(\ref{mode}) reduces to a
harmonic oscillator equation with the solution $\chi \sim e^{i \omega
\eta}$ and the linear (comoving) dispersion relation $\omega = \pm
k$. Equivalently, the dispersion relation expressed in terms of the
proper frequency, $\nu = \omega/a(\eta)$, and wavenumber, $\kappa =
k/a(\eta)$ is simply $\nu = \pm \kappa$. 

The theory is quantized by treating the field as a self-adjoint
operator $\hat \phi$ and imposing the equal-time commutation relations
on the field and its canonically conjugate momenta or, equivalently,
on $\hat a_{\bf k}$ and $\hat a_{\bf k}^\dagger$ \cite{BD84}. In
addition, a vacuum state for $\hat \phi$ must be specified. Since no
unique representation of the vacuum exists in non-stationary
space-times, one usually picks the state that minimizes the
measurement of inflaton ``particles'' at the beginning of inflation by
imposing the adiabatic positive frequency condition at past conformal
infinity ($\eta \to -\infty$) \cite{BD84}. Together with the
normalization provided by the commutation relations, this gives
\cite{BD78}
\begin{equation}
\label{bunchdavies}
\chi(\eta_{\rm i}) = \frac{1}{2}\, \sqrt{\pi \eta_{\rm i}} \, {\cal
H}^{(2)}_{3/2}(k\eta_{\rm i})\,\,, 
\end{equation}
in terms of the Hankel function ${\cal H}^{(2)}$. We can obtain an
estimate for the strength of metric perturbations by evaluating
$|u_{\bf k}|$ at horizon crossing, i.e.~$k = a(\eta) H = -1/\eta$,
yielding the well-known Hawking amplitude: 
\begin{equation}
\label{hawk}
\left(\frac{k^3}{2 \pi^2} |u_{\bf k}|^2 \right)^{1/2} \approx
\frac{H}{2 \pi}\,\,. 
\end{equation}

Our goal is to construct the cosmological analogue of the Unruh model
\cite{U95}, i.e.~a modification of the scalar wave equation that keeps
$\omega$ from exceeding an upper bound. We can then check whether the
prediction for the horizon crossing amplitude deviates from
Eq.~(\ref{hawk}). To this end, we replace the spatial derivative
operator $\nabla$ in Eq.~(\ref{dSKG}) with $F(\nabla, a(\eta))$ where
$F(k,a)$ is an odd, analytic function that has the following
properties:
\begin{equation}
\label{Fprops}
F(k,a) =  \left \{ 
\begin{array}{ll} k \, & {\rm for} \qquad k \ll a\, \kappa_0 \\ 
a\, \kappa_0 \, & {\rm for} \qquad k \gg a\, \kappa_0\,\,,
\end{array}\right.
\end{equation}
thereby explicitly breaking local Lorentz invariance for high
wavenumbers. The preferred frame is the one where $\kappa_0$ is
specified, i.e.~the cosmic rest frame. As discussed, one would
typically expect $\kappa_0 \approx \mpl$. After going through the mode
expansion, the new equation for the mode functions is given by
\begin{equation}
\label{newmode}
\chi'' + \left(F^2(k,a(\eta)) - \frac{a''(\eta)}{a(\eta)}\right)\chi =
0\,\,,
\end{equation}
yielding the modified dispersion relation
\begin{equation}
\label{newdisp}
w  = \pm F(k,a(\eta))
\end{equation}
on subhorizon scales. As promised, the proper frequency is constant,
$\nu = \kappa_0$, far above the proper cutoff scale $\kappa_0$, while
converging to the conventional result far below the cutoff. The
intermediate region around the cutoff may be everything from a smooth
transition over many decades of $k$ to a sharp turnover.

\subsection{Choice of the vacuum state}
\label{initial}

Like in the non-dispersive case, a unique state that minimizes the
probability for particle detection and hence most closely resembles
the standard notion of a vacuum state can be constructed with the help
of an adiabatic expansion \cite{BD84}. To lowest order, an equation of
the form 
\begin{equation}
\label{waveeq}
\chi'' + \omega^2(\eta) \chi = 0
\end{equation}
has the positive frequency WKB solution
\begin{equation}
\label{WKB}
\chi = \frac{1}{\sqrt{2 \omega_+(\eta)}} \exp\left(-i \int
\omega_+(\tilde \eta) d\tilde \eta\right)\,\,,
\end{equation}
 where $\omega_+(\eta)$ is the positive root of $\omega^2$. The
following order correction to $\omega_+$ in Eq.~(\ref{WKB}) is
$O(\omega'/\omega)$.

Eq.~(\ref{Fprops}) shows that the cutoff is exceeded by any mode with
fixed comoving wavenumber $k$ at very early times when $a(\eta) \to
0$, or at any fixed time $\eta$ for very large $k$. For the purpose of
constructing the vacuum state it therefore suffices to analyze
Eq.~(\ref{newmode}) far above the cutoff where it becomes 
\begin{equation}
\label{largek}
\chi'' + \left(a^2 \kappa_0^2 - \frac{a''(\eta)}{a(\eta)}\right)\chi =
0\,\,.
\end{equation}
Introducing the dimensionless parameter
\begin{equation}
\label{sigdef}
\sigma \equiv \frac{\kappa_0}{H} \approx \frac{\mpl}{H}
\end{equation}
characterizing the size of the de Sitter horizon compared to the
cutoff scale, Eq.~(\ref{largek}) can be written in the form of
Eq.~(\ref{waveeq}) with
\begin{equation}
\label{largek2}
\omega^2(\eta) =  \frac{\sigma^2 - 2}{\eta^2}\,\,,
\end{equation}
possessing the exact solutions
\begin{equation}
\label{chisol}
\chi_\pm = {\cal C}_\pm \, \eta^{\frac{1}{2}\left(1 \pm \sqrt{1 -
4(\sigma^2 - 2)}\right)}\,\,.
\end{equation}

The small amplitude of cosmic microwave background fluctuations
indicates that while our current Hubble scale crossed the horizon
during inflation, $\sigma$ was very large, of order $\gtrsim
10^{5}$. Assuming that this condition was true also at the onset of
inflation where the vacuum condition is evaluated, one finds
\begin{equation}
\omega_+ \approx -\frac{\sigma}{\eta}\,\,,
\end{equation} 
with corrections of order $O(\eta^{-2})$ which quickly vanish as $\eta
\to -\infty$. The adiabatic vacuum for $\sigma \gg 1$ is thus given by 
\begin{equation}
\label{largesigvac}
\chi(\eta) = \frac{1}{\sqrt{2 a \kappa_0}} \exp\left(-i \kappa_0
t\right)\,\,,
\end{equation}
where we made the coordinate transformation to FRW time $t$ by
replacing $a\,d\eta$ with $dt$. This solution is also admitted by the
large-$\sigma$ limit of Eq.~(\ref{chisol}). As intuitively expected,
modes far above the  cutoff and far inside the horizon behave like
free harmonic  oscillators with the fixed, $k$-independent proper
frequency $\kappa_0$.

However, some scenarios for inflation invoke an initial phase where
$\sigma \approx 1$ (e.g.,\cite{L84}). In this case,
Eqs.~(\ref{chisol}) and (\ref{largek2}) show that no oscillating
solutions exist as cosmic expansion dominates the field dynamics on
all scales. A particle interpration is unavailable even locally,
making the construction of an adiabatic vacuum with the usual meaning
impossible.

One can instead resort to postulating a phase of slow expansion prior
to the onset of inflation, taking place in an approximately flat
region of space time. Evidently, as long as $a''/a$ is small compared
to $a^2 \kappa_0^2$, Eq.~(\ref{largek}) has the approximate positive
frequency root
\begin{equation}
\omega_+ \approx a \kappa_0\,\,,
\end{equation}
reproducing the adiabatic vacuum solution of the large-$\sigma$ case,
Eq.~(\ref{largesigvac}). The interpretation in terms of harmonic
oscillators with $k$-independent frequencies is, of course, identical.

The authors of Ref.~\cite{MB00} obtain essentially the same vacuum
state by minimizing the energy density of the scalar field. They also
analyze the evolution of the dispersive theory in an alternative
state, the ``instantaneous Minkowski vacuum''. 

\subsection{Comparison of horizon crossing amplitudes}

It remains to be shown that the dispersive theory conforms to the
asymptotic boundary condition of Sec.~(\ref{implications}) for $\sigma
\gg 1$. While the equations of motion, Eqs.~(\ref{mode}) and
(\ref{newmode}), converge if $a \gg k/\kappa_0$ by virtue of
Eq.~(\ref{Fprops}), this 
is not obviously true for the field amplitudes evolving from different
initial conditions.

Subhorizon modes, i.e. those with wavenumbers $k \gg a H$,
approximately obey a harmonic oscillator equation with fixed frequency
$\omega \sim k$ (linear dispersion, Eq.~(\ref{mode})) or time
dependent frequency $\omega \sim F(k,a)$ (nonlinear dispersion,
Eq.~(\ref{newmode})). In the latter case, we will focus our attention on
slowly varying functions $F$ with respect to $a(\eta)$, such that we
can take advantage of the adiabatic invariant $I = A^2\,\omega$ of a
harmonic oscillator obeying $\omega'/\omega \ll \omega$ whose
amplitude is denoted by $A$ (e.g.,\cite{LL76}). The value of $I$ is
set by the initial conditions and remains conserved as long as the
evolution of the field is adiabatic, i.e.~as long as the frequency
change is slow compared with the frequency itself.

According to the initial conditions Eq.~(\ref{bunchdavies}) and
Eq.~(\ref{largesigvac}), $I = 1/2$ initially for both the standard and
the dispersive case. If the physical wavelength for a mode $k$ becomes
larger than the cutoff length while $k$ is still well inside the
horizon, i.e.~if 
\begin{equation}
-\frac{\sigma}{\eta} \gg k \gg -\frac{1}{\eta}\,\,,
\end{equation}
for some $\eta$ the frequencies of both theories converge and the
invariance of $I$ implies the convergence of the
amplitudes. Consequently, the boundary condition of
Sec.~\ref{implications} is satisfied for large $\sigma$, and the
predictions of both theories for the cosmological power spectrum are
identical. This is in agreement with the findings of
Refs.~\cite{MB00,BM00}.

However, if  the transition between the asymptotically constant and
the linear regime is  very smooth, $F(k=aH,a)$ may be significantly
smaller than the horizon crossing frequency in the non-dispersive
case, $\omega = k = aH$, leading to deviations from the standard
result even if $\sigma \gtrsim 10^5$. Although strictly speaking, the
conservation of $I$ is violated as $k/a$ becomes comparable to $H$ due
to cosmological expansion, the effect will be similar in both cases
and we can assume the near equality of $I$. Consequently, the ratio of
the horizon crossing amplitudes of the  dispersive and non-dispersive
theories is approximately
\begin{equation}
\label{ampdiff}
\frac{\langle|\chi|^2\rangle_{\rm disp}}{\langle|\chi|^2\rangle_{\rm
non-disp}} \approx \frac{\omega_{\rm non-disp}(k = aH)}{\omega_{\rm
disp}(k = aH)} = \frac{aH}{F(aH,a)} > 1\,\,,
\end{equation}
giving rise to possible deviations from scale invariance depending on
the actual shape of $F(k,a)$ and the time dependence of $aH$ in a more
realistic model for inflation. This effect becomes important if a
residual quantum gravitational deformation of the dispersion relation is
noticeable at energies approximately 5 orders of magnitude below the
Planck scale. Present experimental bounds do not strongly constrain
this regime; a more quantitative analysis is under way.

In addition to the incomplete (but fully adiabatic) convergence in
case of a very smoothly varying $F(k,a)$, there exists a second
potential source of deviations from the standard theory: if $F(k,a)$
changes too quickly to allow adiabatic adjustment of the modes, the
conservation of $I$ can be violated even on subhorizon scales. The
significance of this effect will again depend on the particular form
chosen for the dispersion relation.

\section{Discussion}
\label{discussion}

In this work, the cosmological analogue of Unruh's dumb hole model is
analyzed in order to address two issues, a purely cosmological one and
a more fundamental problem concerning the origin of Planckian modes in
expanding space times. 

The cosmological question, also investigated in
Refs.~\cite{MB00,BM00}, probes the (in)sensitivity of the predictions
of inflation -- most notably the scale invariance of the spectrum of
cosmological fluctuations -- with regard to changes of physics near
the Planck scale. Even within the highly idealized framework of the
Unruh model the answer is twofold. Assuming that the cosmological
horizon is much larger than the cutoff scale at all times including
the onset of inflation, one finds an adiabatic vacuum state of the
modified theory whose corresponding $k$-mode amplitude adiabatically
evolves toward the standard one. If full convergence is achieved
before the mode crosses the horizon, the standard predictions of
inflation are unchanged (in agreement with
Refs.~\cite{MB00,BM00}). Otherwise, the perturbation amplitude will be
slightly larger than usual, at a level inversely related to the proper
frequency of the mode at horizon crossing (Eq.~\ref{ampdiff}).

If, however, the horizon and cutoff scales are comparable at the
beginning of inflation, the notion of an adiabatic de Sitter vacuum is
invalid as the modes become overdamped at past conformal infinity and
a local particle interpretation ceases to exist. In
Sec.~(\ref{initial}), this problem is formally   solved by demanding
that the de Sitter stage be preceded by a nearly flat and slowly
expanding phase in which the quantum modes attain their ground
state. One can try to justify this choice on the basis of the initial
conditions for inflation but it remains less satisfactory than in the
standard picture. In this scenario, the adiabatic vacuum is identical
to the one in the large-horizon case, and so are the consequences for
the cosmic fluctuation power spectrum.

We also made an attempt to shed light on the following, more
fundamental question, also raised im Ref.~(\cite{J00}):   assuming
that the number of degrees of freedom of the universe is finite due to
a Planck scale cutoff, cosmological expansion inevitably implies that
this number grows in time. What is the mechanism that assigns a
particular state,   e.g.~adiabatic vacuum, to newly born modes? To
avoid conflict with standard model physics, the vacuum state need only
be attained at low wavenumbers and on scales far inside the
cosmological horizon. In certain scenarios for inflation, however,
these two requirements may not be satisfied. The challenge is thus to
find a simple, self-contained model that includes a process for the
creation of modes that converge to the adiabatic vacuum on low
wavenumber, subhorizon scales.

In the analogous black hole problem, an explicit mechanism of mode
creation can be avoided by conversion of ingoing into outgoing modes
\cite{J96}, at least at first sight (see Ref.~\cite{J00}). This is not
possible in the cosmological case. Furthermore, initial data for the
black hole problem can  always be assigned in the asymptotically flat
region far away from the hole, in contrast with cosmology where the
vacuum is usually defined exactly where we need to modify the theory,
namely at $k \to \infty$.

The Unruh dispersion relation employed in this work sidesteps the mode
creation problem by allowing the proper wavelength to become
infinitely small. It is therefore an inadequate model for the more
interesting case in which both frequency and wavelength are  bounded
by some fundamental scale. There, the cosmological ultraviolet problem
forces us to think explicitly about how the vacuum modes of ordinary
quantum field theory emerge from the ``space-time foam'' or a similar
short distance concept. There are a number of complementary ways that
this question may be addressed. Phenomenological quantum gravity in
the spirit of Refs.~\cite{A99,EMN99} might yield some intuition about
how to make contact with current candidate theories for quantum
gravity. ``Fuzzy'' short distance structures, implemented as
space-time uncertainty relations \cite{KM97}, can be used to define
the cutoff in a purely quantum mechanical manner as was successfully
done for Hawking radiation \cite{BGLe99}. Finally, it is conceivable
that the fluid analogy of ``inverse dissipation'', i.e.~the generation
of finite amplitude modes out of thermal (or space-time) noise, will
allow us to gain insight into the kinds of weak nonlinear mode
interactions that manage to stabilize the waves in their vacuum
state. A possible, albeit highly speculative, side effect of such
nonlinearities may be some degree of non-Gaussianity in the
distribution of cosmic microwave background perturbations.

\begin{acknowledgments}
I thank Ted Jacobson and Bob Wald for many helpful discussions and
comments. I also enjoyed  interesting conversations with Achim Kempf,
Will Kinney, Rocky Kolb, and Angela Olinto. I wish to thank the
members of the Enrico Fermi Institute at the University of Chicago for
their hospitality while most of this work was completed.  
\end{acknowledgments}

\bibliography{../../input/bib/early_universe}

\end{document}